\documentclass[prX,showpacs,superscriptaddress,twocolumn]{revtex4}

\usepackage{graphicx}
\usepackage{amsmath}
\usepackage{amsfonts}
\usepackage{amssymb}
\usepackage{epsf}
\usepackage{hyperref}

\newcommand{\fig}[1]{\textsc{Fig.}~\ref{#1}}

\begin{document}
\title{Realization of tunnel barriers for matter waves using spatial gaps}
\author{P. Cheiney}
\affiliation{Universit\'e de Toulouse, UPS, Laboratoire Collisions Agr\'egats R\'eactivit\'e, IRSAMC; F-31062 Toulouse, France}
\affiliation{CNRS, UMR 5589, F-31062 Toulouse, France}
\author{F. Damon}
\affiliation{Universit\'e de Toulouse, UPS, Laboratoire de Physique Th\'eorique, IRSAMC; F-31062 Toulouse, France}
\affiliation{CNRS, UMR 5152, F-31062 Toulouse, France}
\author{G. Condon}
\affiliation{Universit\'e de Toulouse, UPS, Laboratoire Collisions Agr\'egats R\'eactivit\'e, IRSAMC; F-31062 Toulouse, France}
\affiliation{CNRS, UMR 5589, F-31062 Toulouse, France}
\author{B. Georgeot}
\affiliation{Universit\'e de Toulouse, UPS, Laboratoire de Physique Th\'eorique, IRSAMC; F-31062 Toulouse, France}
\affiliation{CNRS, UMR 5152, F-31062 Toulouse, France}
\author{D.~Gu\'ery-Odelin}
\affiliation{Universit\'e de Toulouse, UPS, Laboratoire Collisions Agr\'egats R\'eactivit\'e, IRSAMC; F-31062 Toulouse, France}
\affiliation{CNRS, UMR 5589, F-31062 Toulouse, France}

\date{\today}

\begin{abstract}
We experimentally demonstrate the trapping of a propagating Bose-Einstein Condensate in a Bragg cavity produced by an attractive optical lattice with a smooth envelope. As a consequence of the envelope, the band gaps become position-dependent and act as mirrors of finite and velocity-dependent reflectivity. We directly observe both the oscillations of the wave packet bouncing in the cavity provided by these spatial gaps and the tunneling out for narrow classes of velocity. Synchronization of different classes of velocity can be achieved by proper shaping of the envelope. This technique can generate single or multiple tunnel barriers for matter waves with a tunable transmission probability, equivalent to a standard barrier of submicron size.
\end{abstract}


\maketitle

%\section{Introduction}

The tunneling effect is a cornerstone of quantum mechanics according to which a particle can penetrate and even pass through a classically impenetrable barrier. This behavior results from the wave nature of particles and is at work in many domains of physics including nuclear desintegration \cite{Gamow,PhysRev.33.127}, quantum electronics \cite{PhysRev.109.603,Esaki}, scanning tunneling microscope \cite{STM}, tunnel ionization \cite{Krausz,CCTDGO} and in various superconducting devices \cite{Josephson1962251,fluxt}. 

For cold atoms placed in optical lattices, the tunnel effect controls the coupling between adjacent wells and is therefore an essential parameter to build the band structure and describe the dynamics \cite{MorschRMP,BlochRMP,CCTDGO}. Recent experiments have demonstrated the possibility to control  this coupling dynamically in amplitude and phase \cite{PhysRevLett.95.260404,PhysRevLett.99.220403,PhysRevLett.100.190405,Struck19082011}. 

Realizing a thin enough single barrier enabling one to investigate atom tunneling dynamics remains very challenging. This has been realized so far in at least three different ways to study, for instance, the ac and dc atomic Josephson effect using (i) the combination of an optical lattice and a harmonic potential \cite{PhysRevLett.95.010402}, (ii) a strongly focussed far-off resonance blue-detuned laser \cite{LLS07} and (iii) RF-dressed potentials \cite{PhysRevLett.106.020407}.

The concept of the tunnel effect has been generalized to other kinds of space. The Landau-Zener transition between the energy bands of a lattice can be seen as a tunnel effect in quasi-momentum space \cite{MorschRMP}. Dynamical tunneling has been introduced to describe the tunneling between classically trapped region in a regular phase space \cite{DavisHeller}. Its extension to a partially chaotic phase space, is referred to as chaos-assisted tunneling and has been observed using a deep and strongly modulated optical lattice \cite{SOR01,Phillips}. The tunneling of magnetic flux across a superconducting wire has been recently observed \cite{fluxt}.

In this Letter, we realize a new kind of tunnel barrier in real space using position-dependent band gaps resulting from the smooth envelope of an optical lattice \cite{PhysRevA.60.2312,IacopoE, IacopoR,LMB11}. In this way, we can generate a Bragg cavity for matter waves with effective mirrors of tunable reflectivity, and directly observe the oscillations of a wavepacket (provided by a Bose-Einstein condensate) inside such a cavity along with single tunneling events whenever the packet bounces off these effective mirrors.

We use an attractive lattice with Gaussian envelope resulting from the interference of two off-resonance and red-detuned Gaussian laser beams. An atom in such a lattice experiences a potential $U(x)=-U_0(x)\sin^2\left(\pi x/d\right)$ where $U_0(x)=U_0 \exp\left(-2x^2/w^2\right)$ accounts for the envelope, $U_0$ the maximum lattice depth, $d$ the lattice spacing, and $w$ the Gaussian envelope waist.

 \begin{figure}[!t]
   \begin{center}
      \includegraphics[width=7.5cm]{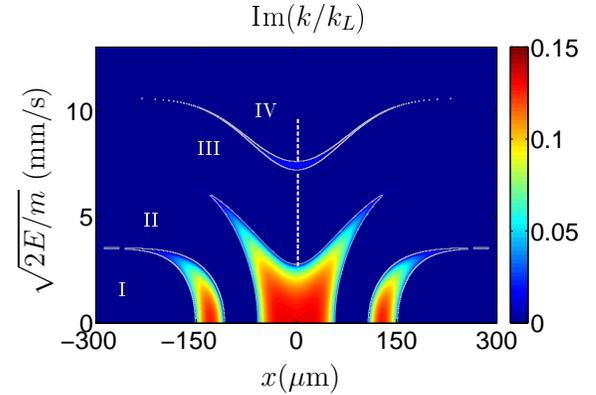}
      \end{center}
\caption{(color online).  Color scale denotes the imaginary part of the Mathieu exponent $k$ with the following lattice parameters: $U_0/E_L=2.5$, $d=650~\rm{nm}$ and $w=145~\rm{\mu m}$. Regions where
$\textrm{Im}(k)\ne0$ correspond to evanescent waves. The roman numbers indicate the band index. The dashed line depicts the initial energy distribution (95 \% of the atoms).}
\label{Fig:bands_sanpera}
\end{figure}

For an infinite optical lattice of constant depth $U_0$, the stationary Schr\"{o}dinger equation is a Mathieu equation \cite{McLachlan,Strang} whose solutions read:
\begin{equation}
\Psi(x;E,U_0)=e^{ik(E)x}u_k(x;E,U_0),
\end{equation}
where $u_k$ is a periodic function of period $d$ and $k$ is the Mathieu exponent. If the energy $E$ corresponds to an allowed band of the Bloch diagram, $k$ is real and corresponds to the pseudo-momentum of a Bloch state. When $E$ lies in a gap, $k$ acquires an imaginary part. The solution of the Mathieu equation is then an evanescent wave that decreases exponentially on a typical length scale $\Delta x_{\rm{tunnel}}=1/\textrm{Im}(k)$ and the real part remains on one edge of the Brillouin zone. 

For a finite optical lattice with a space dependent depth $U(x)$, the same solution locally holds within the assumption of a slowly varying envelope  ($d\ll w$) and yields space-dependent wave vectors $k$. 
In \fig{Fig:bands_sanpera}, we show the imaginary part of the Mathieu exponent as a function of position and square root of the energy for $U_0=2.5E_{\rm L}$ where $E_{\rm{L}}=\hbar^2 k_{\rm{L}}^2/2 m$ with $k_{\rm{L}}=2 \pi/d$ and $m$ is the atom mass. Note that the energy is positive (relative to the continuum far from the attractive lattice) so that a reflection in this context has no classical counterpart and results from a pure quantum effect of matter wave interferences \footnote{The energy $E$ coincides with the kinetic energy $mv_{\rm inc}^2/2$ where $v_{\rm inc}$ is the incoming velocity far from the lattice.} \cite{PhysRevLett.107.230401}.
The regions where $\textrm{Im}(k)\ne 0$ define spatial gaps that separate the different allowed bands. The symmetry of the band structure originates from the symmetry of the Gaussian envelope. %An atom can thus tunnel through a spatial gap if the extent of the gap is smaller than $x_{\rm{tunnel}}$. 

According to this picture, the reflection on a gap corresponds to a Bragg reflection \cite{PhysRevLett.76.4508}, while
tunneling through the barrier provided by the spatial gaps is analogous to a Landau-Zener transition to a different band \cite{Zener01091932}. The same ingredients explain the macroscopic tunneling observed in a vertical lattice \cite{AndersonKasevich}.

\begin{figure}[!t]
   \begin{center}
      \includegraphics[width=6.5cm]{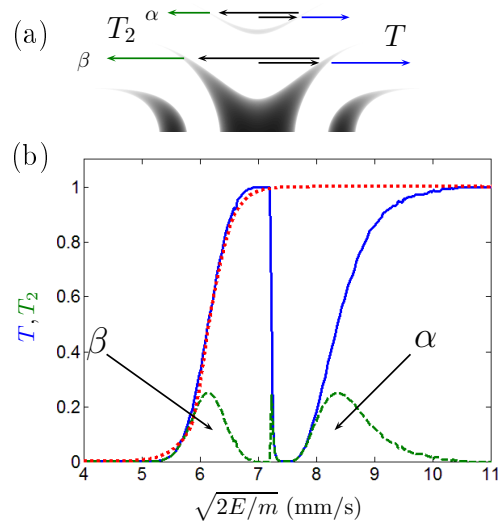}
      \end{center}
\caption{(color online). (a) Schematic of the evolution of the wavepacket inside the lattice (see text). (b) Transmission probability of the wave packet. Solid line (blue): Transmission probability given by Eq.~(\ref{transeq}) through half of the lattice, the regions of reflection correspond to important band gaps. Dotted line (red): Transmission probability through a repulsive Gaussian barrier of variance $\sigma=387~\rm{nm}$. Dashed line (green): Transmission probability to tunnel out of the lattice after one oscillation.  Parameters of the lattice identical to those of \fig{Fig:bands_sanpera}. }
\label{Fig:comp_gauss}
\end{figure}

The transmission probability of a monochromatic wave with energy $E$ through a single barrier can readily be calculated by integrating $\textrm{Im}(k)$
\begin{equation}
T(E)=\exp\left(\int-2\textrm{Im}[k(x,E)]{\rm d}x\right),
\label{transeq}
\end{equation}
and is represented for our parameters as the (blue) solid line in \fig{Fig:comp_gauss}.
The two regions where the transmission probability vanishes correspond to reflections on different spatial gaps. At the edge between regions of transmission and reflection, atoms have a high probability to tunnel through a spatial gap. It is instructive to fit this transmission probability with the one obtained from a repulsive Gaussian barrier. For the spatial gap $\beta$ (see \fig{Fig:comp_gauss}(a)) that is 10 $\mu$m wide, we find the best agreement for a standard deviation of the Gaussian of $\sigma=387~\rm{nm}$ (see red dotted line in \fig{Fig:comp_gauss}(b)). It would be quite challenging to realize such a barrier directly by optical means since it requires to focus a blue-detuned laser of waist $2\sigma$ close to its diffraction limit inside a vacuum chamber.

In our experiment (see below), we initially load the atoms at the center of the lattice with an energy distribution that spreads over the third band and the bottom of the fourth band (see the vertical white dashed line in \fig{Fig:bands_sanpera}) \cite{LMB11}. By energy conservation, the ``trajectory'' of an atom with a well-defined incident velocity remains on a horizontal line in the diagram of \fig{Fig:bands_sanpera}  and may be split on the spatial gaps because of a partial tunneling. Atoms at the bottom of the third band experience a large gap and are thus
reflected with a probability close to one. They bounce back and forth quasi indefinitely. Atoms loaded at the top of the band see
essentially no gap and immediately leave the lattice. Between these two extreme cases, atoms have an intermediate tunneling probability and can leave the trap after one, or several oscillations. The (green) dashed line of \fig{Fig:comp_gauss} represents the
probability $T_2=T(E)(1-T(E))$ to bounce back on the first gap then tunnel out of the cavity at the symmetric position (see \fig{Fig:comp_gauss}(a)). It presents peaks at energies corresponding to tunneling probabilities $T\sim 0.5$. The outcoupling of these atoms can thus be unambiguously attributed to a single tunneling event.

%\section{Experimental setup}
We now turn to our experimental setup. We first produce a $^{87}$Rb Bose-Einstein condensate of typically $3.10^4$ atoms after $3.5~\rm{s}$ of forced evaporation in a crossed dipole trap \cite{PhysRevLett.107.230401}. The trap consists of two red-detuned (1070 nm) laser beams: a
horizontal beam later used as a guide and a dimple beam. During the evaporation, we use the spin-distillation technique to prepare atoms in $\left| F=1, m_F= 1 \right\rangle$ \cite{GCJ09}. Once the Bose-Einstein condensate is formed, the power of the dimple beam is reduced by a factor $\sim20$
over 100~ms to decrease the chemical potential. The condensate is subsequently released in the horizontal guide by switching off abruptly the dimple beam. The  guide confines the transverse degrees of freedom and therefore ensures a quasi one-dimensional dynamics. Our protocol yields a wavepacket with a relatively low velocity dispersion $\Delta v=1.9~\rm{mm/s}$ dictated here by the strength of the interactions combined with our trap decompression. In the course of the propagation,  interactions become rapidly negligible.

The optical lattice is produced at the intersection of two red-detuned laser beams with a wavelength $\lambda_{\rm{L}}=840~\rm{nm}$ and a waist $\tilde{w}=110~\rm{\mu m}$ crossing at an angle $\theta=81^{\circ}$ at the initial position of the wavepacket. The resulting lattice spacing is $d = \lambda_{\rm{L}}/[2 \sin(\theta/2)] \simeq650~\rm{nm}$, and the envelope waist $w=\tilde{w}/\cos(\theta/2)=145~\rm{\mu m}$. The lattice detuning is large enough so that
spontaneous emission is negligible over the duration of the experiment. To calibrate \emph{in situ} the potential depth $U_0$, we use a Kapitza-Dirac diffraction \cite{OMD99}.
%like experiment which consists in switiching on the optical lattice for short amount of time and analyzing the subsequent velocity distribution after a time-of-flight

The wavepacket is set into motion at a mean velocity of 
$\bar{v}=9.4~\rm{mm/s}$ by applying a magnetic field gradient of $14~\rm{G/cm}$ along the guide axis during 4~ms. We then linearly ramp the lattice power up to
$U_0=2.5 E_{\rm{L}}$ in 1~ms. This timescale has been chosen to ensure an adiabatic increase of the lattice depth. This means that the intensity is ramped up in order to keep ${\rm d}\omega/{\rm d}t\ll \omega^2$ where $\omega$ is the trapping frequency at the bottom of one lattice site. According to our numerical simulations, this is already verified for ramping times as short as $100~\rm{\mu s}$. The pseudo-momentum remains constant in the adiabatic loading process. This allows us to determine the final energy distribution of the wavepacket (see the vertical white dashed line in \fig{Fig:bands_sanpera}) \cite{LMB11}. We then let the wavepacket propagate for different times before imaging the atoms \emph{in situ}.

%\section{Experimental results}
\begin{figure}[!t]
   \begin{center}
      \includegraphics[width=7cm]{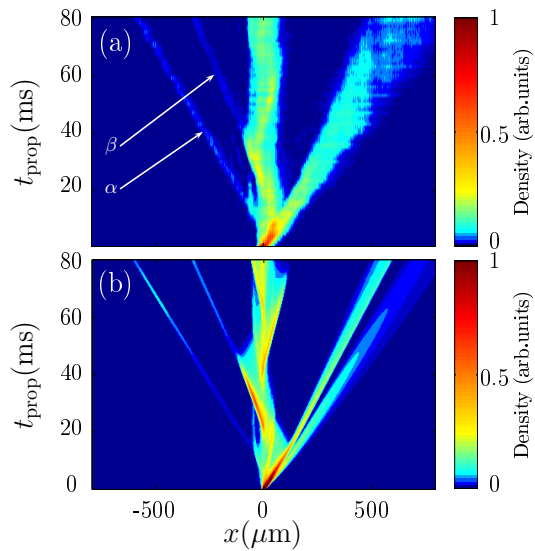}
      \end{center}
\caption{(color online). (a) Measured density distribution of the wavepacket for different propagation times. Each horizontal line
is the average of 4 images integrated along the transverse direction. (b) Direct numerical integration of the Schr\"{o}dinger equation for a
wavepacket whose velocity distribution matches the experimental one.}
\label{Fig:com_exp_simu}
\end{figure}

Figure \ref{Fig:com_exp_simu}(a) shows the measured atomic density along the guide during the propagation. Each horizontal line is the average of four
images integrated along the transverse direction.  In this experiment, all atoms are initially launched toward the right side of the lattice. For the sake of comparison, \fig{Fig:com_exp_simu}(b) is the result of the numerical
integration of the Schr\"{o}dinger equation using a split-Fourier algorithm with a wavepacket whose characteristics match the
experimental ones without any adjustable parameter. %The agreement is very satisfactory. 
Three effects can be noticed: (i) part of the wavepacket immediately leaves the lattice, it corresponds to velocity classes ($6.7 \lesssim v \lesssim 7.2$ and $v \gtrsim9.5$ mm/s) that do not encounter a significant band gap; (ii) a  periodic oscillations inside the lattice can be clearly observed; (iii) in the direction opposite to the initial velocity we observe the emission from the left side of the lattice of two atomic packets  denoted $\alpha$ and $\beta$. They leave the lattice by tunneling through a spatial gap and then propagate freely. Their transmission probabilities have been represented in \fig{Fig:comp_gauss}(b).

The oscillations have a period of approximately 50~ms and appear as regular spines. They are washed out after a few
oscillations. Both of these effects are the consequence of the important initial energy dispersion. The spines structures are
caustics formed by the addition in these regions of the trajectories associated with each energy component. The most energetic atoms travel faster inside the lattice but face the spatial gap at a larger distance from the center for a given spatial gap cavity (see \fig{Fig:bands_sanpera}). For a Gaussian envelope, the second effect turns out to be larger so that the period of oscillation increases with energy. The washing out of the oscillations at long time results from the progressive dephasing of the different energy components.

%\section{Tunneling events}

\begin{figure}[!t]
   \begin{center}
      \includegraphics[width=9cm]{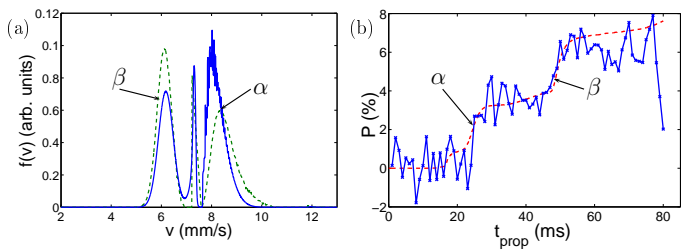}
      \end{center}
\caption{(color online). (a) Calculated velocity distribution of atoms that leave the lattice after one oscillation using the band
model (green dashed line) and the full numerical integration (blue solid line). (b) Proportion of atoms on the left side of the
lattice. Experimental results (blue solid line). Red dashed line is the numerical integration result without any adjustable parameters.}
\label{Fig:tunneled}
\end{figure}

Figure \ref{Fig:tunneled}(b) shows the measured (data of \fig{Fig:com_exp_simu}(a)) and calculated (using the full integration) proportion of atoms on the left side of the lattice at a distance larger than $150~\rm{\mu m}$ from the center as a function of time. It displays two steps  that represent each about 3\% of the total number of atoms and correspond to the two tunneling events observed in \fig{Fig:com_exp_simu}. Atoms on the fourth band generate the first tunnel packet ($\alpha$) in a direction opposite to the launching velocity direction at $t_{\rm{prop}}\simeq25~\rm{ms}$. Atoms in the middle of the third band experience a larger cavity and give rise to the second observed tunnel packet ($\beta$) at $t_{\rm{prop}}\simeq50~\rm{ms}$.

 The measured mean velocities of these narrow wavepackets are $7.9\pm0.1$ and $5.9\pm0.1~\rm{mm/s}$ respectively.
These atoms have performed half an oscillation before leaving the lattice, the atom number and energy distributions can thus be calculated by integrating the probability $T_2$ over the initial energy distribution. We find that the transmissions peaks $\alpha$ and $\beta$ depicted in \fig{Fig:comp_gauss} (dashed line) contain respectively 3.2\% and 3.0\% of the total number of atoms. Figure~\ref{Fig:tunneled}(a) shows the velocity distribution on the left side using the local bands model (green dashed line) and the direct integration (blue solid line). The calculated mean velocity of the two peaks are $5.9$ and $8.3~\rm{mm/s}$ respectively using the band model calculation and $6.1$ and $8.1~\rm{mm/s}$ using the numerical integration in good agreement with the measured values. The two r.m.s velocity dispersions are similar and in the range $\Delta v\simeq250~\rm{\mu m/s}$. This selectivity is as high as the one provided by velocity-selective Raman transitions \cite{KaC92,SOR01,BCG04} and does not require any specific internal state configuration.

\begin{figure}
   \begin{center}
      \includegraphics[width=9cm]{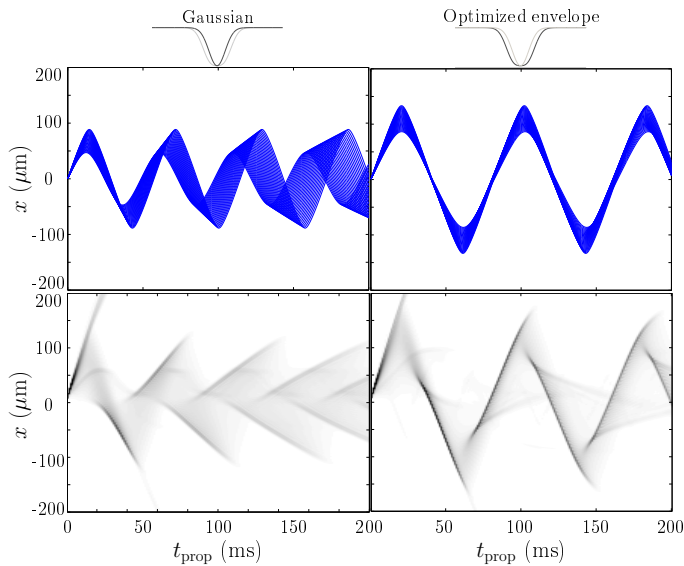}
      \end{center}
\caption{(color online). Envelope engineering. Semiclassical trajectories (upper panel) and full numerical integration (lower panel) of the wavepacket dynamics in (Left) the Gaussian envelope lattice  and in (Right) the lattice with an optimized envelope (see text). Parameters: initial velocity spread $\Delta v=1.2~\rm{mm/s}$, $D=0.75 \,w=112~\rm{\mu m}$.}
\label{Fig:shapping}
\end{figure}

Surprisingly at first sight, the wavepackets that have tunneled out from the spatial gaps $\alpha$ and $\beta$ do not seem to expand significantly even after 80~ms of propagation. Actually, the full numerical integration indicates that the wavepackets are focused at a finite time ($t_{\rm{prop}}\simeq65~\rm{ms}$ for the packet $\alpha$). This is due once again to the dispersion of the oscillation periods: the slow atoms have a shorter period and tunnel before the rapid ones. Outside the lattice this results  in a chirped pulse with the high frequencies at the back. High velocity components then catch up with the slow ones at a finite time. This effect hinders the direct measurement of the velocity dispersion by a time-of-flight.
The dephasing of the different energy components may appear as a limitation of the Bragg cavity device, it enlarges the initial width of the tunneled packets and reduces the number of visible oscillations. Nevertheless, it is possible to circumvent this apparent limitation by keeping the different components in phase with a proper shaping of the envelope.

This shaping consists in adjusting the size of the cavity to compensate exactly for the change in group velocity.
In the following, we demonstrate the optimization of the envelope shape using the \emph{ansatz} 
$U_0(x)=-U_0 \exp(-2x^2/w^2)(1+x^2/D^2)$ where $D$ is a free parameter. Such an envelope can be realized using holographic plates.
 This \emph{ansatz} keeps the symmetry of the Gaussian but has steeper spatial gaps if $D>0$. Because of the caustic effect, it is difficult to define an oscillation period using the numerical integration. Thereby, we have performed this optimization using a semiclassical model \cite{cheiney}. In such an approach, the particle motion on a given local Bloch band $n$ is described by the combined evolution of the wavepacket position and of its mean pseudo-momentum $k$:
\begin{equation}
\dot{x}=\frac{1}{\hbar}\frac{\partial E_n}{\partial k}\hspace{1cm} \textrm{and} \hspace{1cm}\dot{k}=-\frac{1}{\hbar}\frac{\partial E_n}{\partial x}.
\end{equation}
The optimization is performed by cancelling out the first order variation of the oscillation period with energy for trajectories with nearby initial pseudo-momentum \footnote{It is possible to extend the \emph{ansatz} to higher order to cancel out the second
derivative of the oscillation period, however the resulting potentials have a much more complicated shape.}. For our parameters, we find $D=0.75w=112~\rm{\mu m}$. In \fig{Fig:shapping}, we compare the results for a packet of velocity dispersion $\Delta v=1.2~\rm{mm/s}$ for the semiclassical approach and the full numerical integration with and without the optimization. The optimization greatly reduces the blurring of the oscillations, and therefore all velocity components tunnel at the same time, generating a train of nearly identical matter wave pulses i.e. a mode-locked atom laser.

In conclusion, we have demonstrated that spatial gaps resulting from an inhomogeneous envelope of a lattice produce barriers with a probability transmission equivalent to thin real barriers of a few hundreds of nm. They open new perspectives for single tunnel barrier physics including time-modulated tunnel barrier, many-body wavefunctions (such as solitons) tunneling \cite{MNMuga,Sanpera08,Molmer}, Josephson-like experiments \cite{PhysRevLett.106.020407,LLS07}. This system is also of interest for multiple barriers configurations including cavity or Anderson localization investigation in real space \cite{Billy,Inguscio}. It can be readily generalized to higher dimensions and may be used as a test bed for semiclassical approaches of tunneling in 2D \cite{bookS}.

We thank C. Salomon and G. Muga for useful comments.
We acknowledge financial support from the Agence Nationale pour la Recherche, the R\'egion Midi-Pyr\'en\'ees, the university Paul
Sabatier (OMASYC project), the NEXT project ENCOQUAM, the CALMIP supercomputer facility and the Institut Universitaire de France.

%\bibliography{biblio}

\begin{thebibliography}{0}
\expandafter\ifx\csname natexlab\endcsname\relax\def\natexlab#1{#1}\fi
\expandafter\ifx\csname bibnamefont\endcsname\relax
  \def\bibnamefont#1{#1}\fi
\expandafter\ifx\csname bibfnamefont\endcsname\relax
  \def\bibfnamefont#1{#1}\fi
\expandafter\ifx\csname citenamefont\endcsname\relax
  \def\citenamefont#1{#1}\fi
\expandafter\ifx\csname url\endcsname\relax
  \def\url#1{\texttt{#1}}\fi
\expandafter\ifx\csname urlprefix\endcsname\relax\def\urlprefix{URL }\fi
\providecommand{\bibinfo}[2]{#2}
\providecommand{\eprint}[2][]{\url{#2}}

\end{thebibliography}


\begin{thebibliography}{42}
\expandafter\ifx\csname natexlab\endcsname\relax\def\natexlab#1{#1}\fi
\expandafter\ifx\csname bibnamefont\endcsname\relax
  \def\bibnamefont#1{#1}\fi
\expandafter\ifx\csname bibfnamefont\endcsname\relax
  \def\bibfnamefont#1{#1}\fi
\expandafter\ifx\csname citenamefont\endcsname\relax
  \def\citenamefont#1{#1}\fi
\expandafter\ifx\csname url\endcsname\relax
  \def\url#1{\texttt{#1}}\fi
\expandafter\ifx\csname urlprefix\endcsname\relax\def\urlprefix{URL }\fi
\providecommand{\bibinfo}[2]{#2}
\providecommand{\eprint}[2][]{\url{#2}}

\bibitem[{\citenamefont{Gamow}(1928)}]{Gamow}
\bibinfo{author}{\bibfnamefont{G.}~\bibnamefont{Gamow}},
  \bibinfo{journal}{Zeitschrift f\"ur Physik} \textbf{\bibinfo{volume}{51}},
  \bibinfo{pages}{204} (\bibinfo{year}{1928}).

\bibitem[{\citenamefont{Gurney and Condon}(1929)}]{PhysRev.33.127}
\bibinfo{author}{\bibfnamefont{R.~W.} \bibnamefont{Gurney}} \bibnamefont{and}
  \bibinfo{author}{\bibfnamefont{E.~U.} \bibnamefont{Condon}},
  \bibinfo{journal}{Phys. Rev.} \textbf{\bibinfo{volume}{33}},
  \bibinfo{pages}{127} (\bibinfo{year}{1929}).

\bibitem[{\citenamefont{Esaki}(1958)}]{PhysRev.109.603}
\bibinfo{author}{\bibfnamefont{L.}~\bibnamefont{Esaki}},
  \bibinfo{journal}{Phys. Rev.} \textbf{\bibinfo{volume}{109}},
  \bibinfo{pages}{603} (\bibinfo{year}{1958}).

\bibitem[{\citenamefont{Esaki}(1974)}]{Esaki}
\bibinfo{author}{\bibfnamefont{L.}~\bibnamefont{Esaki}}, \bibinfo{journal}{Rev.
  Mod. Phys.} \textbf{\bibinfo{volume}{46}}, \bibinfo{pages}{237}
  (\bibinfo{year}{1974}).

\bibitem[{\citenamefont{Binnig and Rohrer}(1987)}]{STM}
\bibinfo{author}{\bibfnamefont{G.}~\bibnamefont{Binnig}} \bibnamefont{and}
  \bibinfo{author}{\bibfnamefont{H.}~\bibnamefont{Rohrer}},
  \bibinfo{journal}{Rev. Mod. Phys.} \textbf{\bibinfo{volume}{59}},
  \bibinfo{pages}{615} (\bibinfo{year}{1987}).

\bibitem[{\citenamefont{Krausz and Ivanov}(2009)}]{Krausz}
\bibinfo{author}{\bibfnamefont{F.}~\bibnamefont{Krausz}} \bibnamefont{and}
  \bibinfo{author}{\bibfnamefont{M.}~\bibnamefont{Ivanov}},
  \bibinfo{journal}{Rev. Mod. Phys.} \textbf{\bibinfo{volume}{81}},
  \bibinfo{pages}{163} (\bibinfo{year}{2009}).

\bibitem[{\citenamefont{Cohen-Tannoudji and Gu\'ery-Odelin}(2011)}]{CCTDGO}
\bibinfo{author}{\bibfnamefont{C.}~\bibnamefont{Cohen-Tannoudji}}
  \bibnamefont{and}
  \bibinfo{author}{\bibfnamefont{D.}~\bibnamefont{Gu\'ery-Odelin}},
  \emph{\bibinfo{title}{Advances in atomic physics, an overview}}
  (\bibinfo{publisher}{World Scientific, Singapore}, \bibinfo{year}{2011}).

\bibitem[{\citenamefont{Josephson}(1962)}]{Josephson1962251}
\bibinfo{author}{\bibfnamefont{B.}~\bibnamefont{Josephson}},
  \bibinfo{journal}{Physics Letters} \textbf{\bibinfo{volume}{1}},
  \bibinfo{pages}{251 } (\bibinfo{year}{1962}).

\bibitem[{\citenamefont{Astafiev et~al.}(2012)\citenamefont{Astafiev, Ioffe,
  Kafanov, Pashkin, Arutyunov, Shahar, Cohen, and Tsai}}]{fluxt}
\bibinfo{author}{\bibfnamefont{O.~V.} \bibnamefont{Astafiev}},
  \bibinfo{author}{\bibfnamefont{L.~B.} \bibnamefont{Ioffe}},
  \bibinfo{author}{\bibfnamefont{S.}~\bibnamefont{Kafanov}},
  \bibinfo{author}{\bibfnamefont{Y.~A.} \bibnamefont{Pashkin}},
  \bibinfo{author}{\bibfnamefont{K.~Y.} \bibnamefont{Arutyunov}},
  \bibinfo{author}{\bibfnamefont{D.}~\bibnamefont{Shahar}},
  \bibinfo{author}{\bibfnamefont{O.}~\bibnamefont{Cohen}}, \bibnamefont{and}
  \bibinfo{author}{\bibfnamefont{J.~S.} \bibnamefont{Tsai}},
  \bibinfo{journal}{Nature} \textbf{\bibinfo{volume}{484}},
  \bibinfo{pages}{355} (\bibinfo{year}{2012}).

\bibitem[{\citenamefont{Morsch and Oberthaler}(2006)}]{MorschRMP}
\bibinfo{author}{\bibfnamefont{O.}~\bibnamefont{Morsch}} \bibnamefont{and}
  \bibinfo{author}{\bibfnamefont{M.}~\bibnamefont{Oberthaler}},
  \bibinfo{journal}{Rev. Mod. Phys.} \textbf{\bibinfo{volume}{78}},
  \bibinfo{pages}{179} (\bibinfo{year}{2006}).

\bibitem[{\citenamefont{Bloch et~al.}(2008)\citenamefont{Bloch, Dalibard, and
  Zwerger}}]{BlochRMP}
\bibinfo{author}{\bibfnamefont{I.}~\bibnamefont{Bloch}},
  \bibinfo{author}{\bibfnamefont{J.}~\bibnamefont{Dalibard}}, \bibnamefont{and}
  \bibinfo{author}{\bibfnamefont{W.}~\bibnamefont{Zwerger}},
  \bibinfo{journal}{Rev. Mod. Phys.} \textbf{\bibinfo{volume}{80}},
  \bibinfo{pages}{885} (\bibinfo{year}{2008}).

\bibitem[{\citenamefont{Eckardt et~al.}(2005)\citenamefont{Eckardt, Weiss, and
  Holthaus}}]{PhysRevLett.95.260404}
\bibinfo{author}{\bibfnamefont{A.}~\bibnamefont{Eckardt}},
  \bibinfo{author}{\bibfnamefont{C.}~\bibnamefont{Weiss}}, \bibnamefont{and}
  \bibinfo{author}{\bibfnamefont{M.}~\bibnamefont{Holthaus}},
  \bibinfo{journal}{Phys. Rev. Lett.} \textbf{\bibinfo{volume}{95}},
  \bibinfo{pages}{260404} (\bibinfo{year}{2005}).

\bibitem[{\citenamefont{Lignier et~al.}(2007)\citenamefont{Lignier, Sias,
  Ciampini, Singh, Zenesini, Morsch, and Arimondo}}]{PhysRevLett.99.220403}
\bibinfo{author}{\bibfnamefont{H.}~\bibnamefont{Lignier}},
  \bibinfo{author}{\bibfnamefont{C.}~\bibnamefont{Sias}},
  \bibinfo{author}{\bibfnamefont{D.}~\bibnamefont{Ciampini}},
  \bibinfo{author}{\bibfnamefont{Y.}~\bibnamefont{Singh}},
  \bibinfo{author}{\bibfnamefont{A.}~\bibnamefont{Zenesini}},
  \bibinfo{author}{\bibfnamefont{O.}~\bibnamefont{Morsch}}, \bibnamefont{and}
  \bibinfo{author}{\bibfnamefont{E.}~\bibnamefont{Arimondo}},
  \bibinfo{journal}{Phys. Rev. Lett.} \textbf{\bibinfo{volume}{99}},
  \bibinfo{pages}{220403} (\bibinfo{year}{2007}).

\bibitem[{\citenamefont{Kierig et~al.}(2008)\citenamefont{Kierig,
  Schnorrberger, Schietinger, Tomkovic, and
  Oberthaler}}]{PhysRevLett.100.190405}
\bibinfo{author}{\bibfnamefont{E.}~\bibnamefont{Kierig}},
  \bibinfo{author}{\bibfnamefont{U.}~\bibnamefont{Schnorrberger}},
  \bibinfo{author}{\bibfnamefont{A.}~\bibnamefont{Schietinger}},
  \bibinfo{author}{\bibfnamefont{J.}~\bibnamefont{Tomkovic}}, \bibnamefont{and}
  \bibinfo{author}{\bibfnamefont{M.~K.} \bibnamefont{Oberthaler}},
  \bibinfo{journal}{Phys. Rev. Lett.} \textbf{\bibinfo{volume}{100}},
  \bibinfo{pages}{190405} (\bibinfo{year}{2008}).

\bibitem[{\citenamefont{Struck et~al.}(2011)\citenamefont{Struck, Ölschläger,
  Le~Targat, Soltan-Panahi, Eckardt, Lewenstein, Windpassinger, and
  Sengstock}}]{Struck19082011}
\bibinfo{author}{\bibfnamefont{J.}~\bibnamefont{Struck}},
  \bibinfo{author}{\bibfnamefont{C.}~\bibnamefont{Ölschläger}},
  \bibinfo{author}{\bibfnamefont{R.}~\bibnamefont{Le~Targat}},
  \bibinfo{author}{\bibfnamefont{P.}~\bibnamefont{Soltan-Panahi}},
  \bibinfo{author}{\bibfnamefont{A.}~\bibnamefont{Eckardt}},
  \bibinfo{author}{\bibfnamefont{M.}~\bibnamefont{Lewenstein}},
  \bibinfo{author}{\bibfnamefont{P.}~\bibnamefont{Windpassinger}},
  \bibnamefont{and}
  \bibinfo{author}{\bibfnamefont{K.}~\bibnamefont{Sengstock}},
  \bibinfo{journal}{Science} \textbf{\bibinfo{volume}{333}},
  \bibinfo{pages}{996} (\bibinfo{year}{2011}).

\bibitem[{\citenamefont{Albiez et~al.}(2005)\citenamefont{Albiez, Gati,
  F\"olling, Hunsmann, Cristiani, and Oberthaler}}]{PhysRevLett.95.010402}
\bibinfo{author}{\bibfnamefont{M.}~\bibnamefont{Albiez}},
  \bibinfo{author}{\bibfnamefont{R.}~\bibnamefont{Gati}},
  \bibinfo{author}{\bibfnamefont{J.}~\bibnamefont{F\"olling}},
  \bibinfo{author}{\bibfnamefont{S.}~\bibnamefont{Hunsmann}},
  \bibinfo{author}{\bibfnamefont{M.}~\bibnamefont{Cristiani}},
  \bibnamefont{and} \bibinfo{author}{\bibfnamefont{M.~K.}
  \bibnamefont{Oberthaler}}, \bibinfo{journal}{Phys. Rev. Lett.}
  \textbf{\bibinfo{volume}{95}}, \bibinfo{pages}{010402}
  (\bibinfo{year}{2005}).

\bibitem[{\citenamefont{Levy et~al.}(2007)\citenamefont{Levy, Lahoud, Shomroni,
  and Steinhauer}}]{LLS07}
\bibinfo{author}{\bibfnamefont{S.}~\bibnamefont{Levy}},
  \bibinfo{author}{\bibfnamefont{E.}~\bibnamefont{Lahoud}},
  \bibinfo{author}{\bibfnamefont{I.}~\bibnamefont{Shomroni}}, \bibnamefont{and}
  \bibinfo{author}{\bibfnamefont{J.}~\bibnamefont{Steinhauer}},
  \bibinfo{journal}{Nature} \textbf{\bibinfo{volume}{449}},
  \bibinfo{pages}{579} (\bibinfo{year}{2007}).

\bibitem[{\citenamefont{Betz et~al.}(2011)\citenamefont{Betz, Manz, B\"ucker,
  Berrada, Koller, Kazakov, Mazets, Stimming, Perrin, Schumm
  et~al.}}]{PhysRevLett.106.020407}
\bibinfo{author}{\bibfnamefont{T.}~\bibnamefont{Betz}},
  \bibinfo{author}{\bibfnamefont{S.}~\bibnamefont{Manz}},
  \bibinfo{author}{\bibfnamefont{R.}~\bibnamefont{B\"ucker}},
  \bibinfo{author}{\bibfnamefont{T.}~\bibnamefont{Berrada}},
  \bibinfo{author}{\bibfnamefont{C.}~\bibnamefont{Koller}},
  \bibinfo{author}{\bibfnamefont{G.}~\bibnamefont{Kazakov}},
  \bibinfo{author}{\bibfnamefont{I.~E.} \bibnamefont{Mazets}},
  \bibinfo{author}{\bibfnamefont{H.-P.} \bibnamefont{Stimming}},
  \bibinfo{author}{\bibfnamefont{A.}~\bibnamefont{Perrin}},
  \bibinfo{author}{\bibfnamefont{T.}~\bibnamefont{Schumm}},
  \bibnamefont{et~al.}, \bibinfo{journal}{Phys. Rev. Lett.}
  \textbf{\bibinfo{volume}{106}}, \bibinfo{pages}{020407}
  (\bibinfo{year}{2011}).

\bibitem[{\citenamefont{Davis and Heller}(1981)}]{DavisHeller}
\bibinfo{author}{\bibfnamefont{M.~J.} \bibnamefont{Davis}} \bibnamefont{and}
  \bibinfo{author}{\bibfnamefont{E.~J.} \bibnamefont{Heller}},
  \bibinfo{journal}{J. Chem. Phys.} \textbf{\bibinfo{volume}{75}},
  \bibinfo{pages}{246} (\bibinfo{year}{1981}).

\bibitem[{\citenamefont{Steck et~al.}(2001)\citenamefont{Steck, Oskay, and
  Raizen}}]{SOR01}
\bibinfo{author}{\bibfnamefont{D.~A.} \bibnamefont{Steck}},
  \bibinfo{author}{\bibfnamefont{W.~H.} \bibnamefont{Oskay}}, \bibnamefont{and}
  \bibinfo{author}{\bibfnamefont{M.~G.} \bibnamefont{Raizen}},
  \bibinfo{journal}{Science} \textbf{\bibinfo{volume}{293}},
  \bibinfo{pages}{274} (\bibinfo{year}{2001}).

\bibitem[{\citenamefont{Haffner et~al.}(2001)\citenamefont{Haffner, Browaeys,
  Heckenberg, Helmerson, McKenzie, Milburn, Phillips, Rolston,
  Rubinsztein-Dunlop, and Upcroft}}]{Phillips}
\bibinfo{author}{\bibfnamefont{H.}~\bibnamefont{Haffner}},
  \bibinfo{author}{\bibfnamefont{A.}~\bibnamefont{Browaeys}},
  \bibinfo{author}{\bibfnamefont{N.~R.} \bibnamefont{Heckenberg}},
  \bibinfo{author}{\bibfnamefont{K.}~\bibnamefont{Helmerson}},
  \bibinfo{author}{\bibfnamefont{C.}~\bibnamefont{McKenzie}},
  \bibinfo{author}{\bibfnamefont{G.~J.} \bibnamefont{Milburn}},
  \bibinfo{author}{\bibfnamefont{W.~D.} \bibnamefont{Phillips}},
  \bibinfo{author}{\bibfnamefont{S.~L.} \bibnamefont{Rolston}},
  \bibinfo{author}{\bibfnamefont{H.}~\bibnamefont{Rubinsztein-Dunlop}},
  \bibnamefont{and} \bibinfo{author}{\bibfnamefont{B.}~\bibnamefont{Upcroft}},
  \bibinfo{journal}{Nature} \textbf{\bibinfo{volume}{412}}, \bibinfo{pages}{52}
  (\bibinfo{year}{2001}).

\bibitem[{\citenamefont{Santos and Roso}(1999)}]{PhysRevA.60.2312}
\bibinfo{author}{\bibfnamefont{L.}~\bibnamefont{Santos}} \bibnamefont{and}
  \bibinfo{author}{\bibfnamefont{L.}~\bibnamefont{Roso}},
  \bibinfo{journal}{Phys. Rev. A} \textbf{\bibinfo{volume}{60}},
  \bibinfo{pages}{2312} (\bibinfo{year}{1999}).

\bibitem[{\citenamefont{Carusotto et~al.}(2002)\citenamefont{Carusotto,
  Embriaco, and La~Rocca}}]{IacopoE}
\bibinfo{author}{\bibfnamefont{I.}~\bibnamefont{Carusotto}},
  \bibinfo{author}{\bibfnamefont{D.}~\bibnamefont{Embriaco}}, \bibnamefont{and}
  \bibinfo{author}{\bibfnamefont{G.~C.} \bibnamefont{La~Rocca}},
  \bibinfo{journal}{Phys. Rev. A} \textbf{\bibinfo{volume}{65}},
  \bibinfo{pages}{053611} (\bibinfo{year}{2002}).

\bibitem[{\citenamefont{Carusotto and La~Rocca}(2000)}]{IacopoR}
\bibinfo{author}{\bibfnamefont{I.}~\bibnamefont{Carusotto}} \bibnamefont{and}
  \bibinfo{author}{\bibfnamefont{G.~C.} \bibnamefont{La~Rocca}},
  \bibinfo{journal}{Phys. Rev. Lett.} \textbf{\bibinfo{volume}{84}},
  \bibinfo{pages}{399} (\bibinfo{year}{2000}).

\bibitem[{\citenamefont{Lauber et~al.}(2011)\citenamefont{Lauber, Massignan,
  Birkl, and Sanpera}}]{LMB11}
\bibinfo{author}{\bibfnamefont{T.}~\bibnamefont{Lauber}},
  \bibinfo{author}{\bibfnamefont{P.}~\bibnamefont{Massignan}},
  \bibinfo{author}{\bibfnamefont{G.}~\bibnamefont{Birkl}}, \bibnamefont{and}
  \bibinfo{author}{\bibfnamefont{A.}~\bibnamefont{Sanpera}},
  \bibinfo{journal}{Journal of Physics B: Atomic, Molecular and Optical
  Physics} \textbf{\bibinfo{volume}{44}}, \bibinfo{pages}{065301}
  (\bibinfo{year}{2011}).

\bibitem[{\citenamefont{Magnus and Winkler}(1947)}]{McLachlan}
\bibinfo{author}{\bibfnamefont{W.}~\bibnamefont{Magnus}} \bibnamefont{and}
  \bibinfo{author}{\bibfnamefont{S.}~\bibnamefont{Winkler}},
  \emph{\bibinfo{title}{Hill's equation}} (\bibinfo{publisher}{John Wiley and
  Sons, New York}, \bibinfo{year}{1947}).

\bibitem[{\citenamefont{Str\"ang}(2005)}]{Strang}
\bibinfo{author}{\bibfnamefont{J.~E.} \bibnamefont{Str\"ang}},
  \bibinfo{journal}{Acad. Roy. Belg. Bull. Cl. Sci.}
  \textbf{\bibinfo{volume}{6}}, \bibinfo{pages}{269} (\bibinfo{year}{2005}).

\bibitem[{\citenamefont{Fabre et~al.}(2011)\citenamefont{Fabre, Cheiney,
  Gattobigio, Vermersch, Faure, Mathevet, Lahaye, and
  Gu\'ery-Odelin}}]{PhysRevLett.107.230401}
\bibinfo{author}{\bibfnamefont{C.~M.} \bibnamefont{Fabre}},
  \bibinfo{author}{\bibfnamefont{P.}~\bibnamefont{Cheiney}},
  \bibinfo{author}{\bibfnamefont{G.~L.} \bibnamefont{Gattobigio}},
  \bibinfo{author}{\bibfnamefont{F.}~\bibnamefont{Vermersch}},
  \bibinfo{author}{\bibfnamefont{S.}~\bibnamefont{Faure}},
  \bibinfo{author}{\bibfnamefont{R.}~\bibnamefont{Mathevet}},
  \bibinfo{author}{\bibfnamefont{T.}~\bibnamefont{Lahaye}}, \bibnamefont{and}
  \bibinfo{author}{\bibfnamefont{D.}~\bibnamefont{Gu\'ery-Odelin}},
  \bibinfo{journal}{Phys. Rev. Lett.} \textbf{\bibinfo{volume}{107}},
  \bibinfo{pages}{230401} (\bibinfo{year}{2011}).

\bibitem[{\citenamefont{Ben~Dahan et~al.}(1996)\citenamefont{Ben~Dahan, Peik,
  Reichel, Castin, and Salomon}}]{PhysRevLett.76.4508}
\bibinfo{author}{\bibfnamefont{M.}~\bibnamefont{Ben~Dahan}},
  \bibinfo{author}{\bibfnamefont{E.}~\bibnamefont{Peik}},
  \bibinfo{author}{\bibfnamefont{J.}~\bibnamefont{Reichel}},
  \bibinfo{author}{\bibfnamefont{Y.}~\bibnamefont{Castin}}, \bibnamefont{and}
  \bibinfo{author}{\bibfnamefont{C.}~\bibnamefont{Salomon}},
  \bibinfo{journal}{Phys. Rev. Lett.} \textbf{\bibinfo{volume}{76}},
  \bibinfo{pages}{4508} (\bibinfo{year}{1996}).

\bibitem[{\citenamefont{Zener}(1932)}]{Zener01091932}
\bibinfo{author}{\bibfnamefont{C.}~\bibnamefont{Zener}},
  \bibinfo{journal}{Proceedings of the Royal Society of London. Series A}
  \textbf{\bibinfo{volume}{137}}, \bibinfo{pages}{696} (\bibinfo{year}{1932}).

\bibitem[{\citenamefont{Anderson and Kasevich}(1998)}]{AndersonKasevich}
\bibinfo{author}{\bibfnamefont{B.~P.} \bibnamefont{Anderson}} \bibnamefont{and}
  \bibinfo{author}{\bibfnamefont{M.~A.} \bibnamefont{Kasevich}},
  \bibinfo{journal}{Science} \textbf{\bibinfo{volume}{282}},
  \bibinfo{pages}{1686} (\bibinfo{year}{1998}).

\bibitem[{\citenamefont{Gattobigio et~al.}(2009)\citenamefont{Gattobigio,
  Couvert, Jeppesen, Mathevet, and Gu\'ery-Odelin}}]{GCJ09}
\bibinfo{author}{\bibfnamefont{G.~L.} \bibnamefont{Gattobigio}},
  \bibinfo{author}{\bibfnamefont{A.}~\bibnamefont{Couvert}},
  \bibinfo{author}{\bibfnamefont{M.}~\bibnamefont{Jeppesen}},
  \bibinfo{author}{\bibfnamefont{R.}~\bibnamefont{Mathevet}}, \bibnamefont{and}
  \bibinfo{author}{\bibfnamefont{D.}~\bibnamefont{Gu\'ery-Odelin}},
  \bibinfo{journal}{Phys. Rev. A} \textbf{\bibinfo{volume}{80}},
  \bibinfo{pages}{041605} (\bibinfo{year}{2009}).

\bibitem[{\citenamefont{Ovchinnikov et~al.}(1999)\citenamefont{Ovchinnikov,
  M\"uller, Doery, Vredenbregt, Helmerson, Rolston, and Phillips}}]{OMD99}
\bibinfo{author}{\bibfnamefont{Y.~B.} \bibnamefont{Ovchinnikov}},
  \bibinfo{author}{\bibfnamefont{J.~H.} \bibnamefont{M\"uller}},
  \bibinfo{author}{\bibfnamefont{M.~R.} \bibnamefont{Doery}},
  \bibinfo{author}{\bibfnamefont{E.~J.~D.} \bibnamefont{Vredenbregt}},
  \bibinfo{author}{\bibfnamefont{K.}~\bibnamefont{Helmerson}},
  \bibinfo{author}{\bibfnamefont{S.~L.} \bibnamefont{Rolston}},
  \bibnamefont{and} \bibinfo{author}{\bibfnamefont{W.~D.}
  \bibnamefont{Phillips}}, \bibinfo{journal}{Phys. Rev. Lett.}
  \textbf{\bibinfo{volume}{83}}, \bibinfo{pages}{284} (\bibinfo{year}{1999}).

\bibitem[{\citenamefont{Kasevich and Chu}(1992)}]{KaC92}
\bibinfo{author}{\bibfnamefont{M.}~\bibnamefont{Kasevich}} \bibnamefont{and}
  \bibinfo{author}{\bibfnamefont{S.}~\bibnamefont{Chu}},
  \bibinfo{journal}{Phys. Rev. Lett.} \textbf{\bibinfo{volume}{69}},
  \bibinfo{pages}{1741} (\bibinfo{year}{1992}).

\bibitem[{\citenamefont{Battesti et~al.}(2004)\citenamefont{Battesti, Clad\'e,
  Guellati-Kh\'elifa, Schwob, Gr\'emaud, Nez, Julien, and Biraben}}]{BCG04}
\bibinfo{author}{\bibfnamefont{R.}~\bibnamefont{Battesti}},
  \bibinfo{author}{\bibfnamefont{P.}~\bibnamefont{Clad\'e}},
  \bibinfo{author}{\bibfnamefont{S.}~\bibnamefont{Guellati-Kh\'elifa}},
  \bibinfo{author}{\bibfnamefont{C.}~\bibnamefont{Schwob}},
  \bibinfo{author}{\bibfnamefont{B.}~\bibnamefont{Gr\'emaud}},
  \bibinfo{author}{\bibfnamefont{F.}~\bibnamefont{Nez}},
  \bibinfo{author}{\bibfnamefont{L.}~\bibnamefont{Julien}}, \bibnamefont{and}
  \bibinfo{author}{\bibfnamefont{F.}~\bibnamefont{Biraben}},
  \bibinfo{journal}{Phys. Rev. Lett.} \textbf{\bibinfo{volume}{92}},
  \bibinfo{pages}{253001} (\bibinfo{year}{2004}).

\bibitem[{\citenamefont{Cheiney et~al.}(2013)\citenamefont{Cheiney, Fabre,
  Vermersch, Gattobigio, Mathevet, Lahaye, and Gu\'ery-Odelin}}]{cheiney}
\bibinfo{author}{\bibfnamefont{P.}~\bibnamefont{Cheiney}},
  \bibinfo{author}{\bibfnamefont{C.~M.} \bibnamefont{Fabre}},
  \bibinfo{author}{\bibfnamefont{F.}~\bibnamefont{Vermersch}},
  \bibinfo{author}{\bibfnamefont{G.~L.} \bibnamefont{Gattobigio}},
  \bibinfo{author}{\bibfnamefont{R.}~\bibnamefont{Mathevet}},
  \bibinfo{author}{\bibfnamefont{T.}~\bibnamefont{Lahaye}}, \bibnamefont{and}
  \bibinfo{author}{\bibfnamefont{D.}~\bibnamefont{Gu\'ery-Odelin}},
  \bibinfo{journal}{Phys. Rev. A} \textbf{\bibinfo{volume}{87}},
  \bibinfo{pages}{013623} (\bibinfo{year}{2013}).

\bibitem[{\citenamefont{del Campo et~al.}(2006)\citenamefont{del Campo,
  Delgado, Garcia-Calderon, Muga, and Raizen}}]{MNMuga}
\bibinfo{author}{\bibfnamefont{A.}~\bibnamefont{del Campo}},
  \bibinfo{author}{\bibfnamefont{F.}~\bibnamefont{Delgado}},
  \bibinfo{author}{\bibfnamefont{G.}~\bibnamefont{Garcia-Calderon}},
  \bibinfo{author}{\bibfnamefont{J.~G.} \bibnamefont{Muga}}, \bibnamefont{and}
  \bibinfo{author}{\bibfnamefont{M.~G.} \bibnamefont{Raizen}},
  \bibinfo{journal}{Phys. Rev. A} \textbf{\bibinfo{volume}{74}},
  \bibinfo{pages}{013605} (\bibinfo{year}{2006}).

\bibitem[{\citenamefont{Ahufinger et~al.}(2008)\citenamefont{Ahufinger,
  Malomed, Birkl, Corbal\'an, and Sanpera}}]{Sanpera08}
\bibinfo{author}{\bibfnamefont{V.}~\bibnamefont{Ahufinger}},
  \bibinfo{author}{\bibfnamefont{B.~A.} \bibnamefont{Malomed}},
  \bibinfo{author}{\bibfnamefont{G.}~\bibnamefont{Birkl}},
  \bibinfo{author}{\bibfnamefont{R.}~\bibnamefont{Corbal\'an}},
  \bibnamefont{and} \bibinfo{author}{\bibfnamefont{A.}~\bibnamefont{Sanpera}},
  \bibinfo{journal}{Phys. Rev. A} \textbf{\bibinfo{volume}{78}},
  \bibinfo{pages}{013608} (\bibinfo{year}{2008}).

\bibitem[{\citenamefont{Hansen et~al.}(2012)\citenamefont{Hansen, Nygaard, and
  Molmer}}]{Molmer}
\bibinfo{author}{\bibfnamefont{S.~D.} \bibnamefont{Hansen}},
  \bibinfo{author}{\bibfnamefont{N.}~\bibnamefont{Nygaard}}, \bibnamefont{and}
  \bibinfo{author}{\bibfnamefont{K.}~\bibnamefont{Molmer}},
  \bibinfo{journal}{arXiv:1210.1681}  (\bibinfo{year}{2012}).

\bibitem[{\citenamefont{Billy et~al.}(2008)\citenamefont{Billy, Josse, Zuo,
  Bernard, Hambrecht, Lugan, Clement, Sanchez-Palencia, Bouyer, and
  Aspect}}]{Billy}
\bibinfo{author}{\bibfnamefont{J.}~\bibnamefont{Billy}},
  \bibinfo{author}{\bibfnamefont{V.}~\bibnamefont{Josse}},
  \bibinfo{author}{\bibfnamefont{Z.}~\bibnamefont{Zuo}},
  \bibinfo{author}{\bibfnamefont{A.}~\bibnamefont{Bernard}},
  \bibinfo{author}{\bibfnamefont{B.}~\bibnamefont{Hambrecht}},
  \bibinfo{author}{\bibfnamefont{P.}~\bibnamefont{Lugan}},
  \bibinfo{author}{\bibfnamefont{D.}~\bibnamefont{Clement}},
  \bibinfo{author}{\bibfnamefont{L.}~\bibnamefont{Sanchez-Palencia}},
  \bibinfo{author}{\bibfnamefont{P.}~\bibnamefont{Bouyer}}, \bibnamefont{and}
  \bibinfo{author}{\bibfnamefont{A.}~\bibnamefont{Aspect}},
  \bibinfo{journal}{Nature} \textbf{\bibinfo{volume}{453}},
  \bibinfo{pages}{891} (\bibinfo{year}{2008}).

\bibitem[{\citenamefont{Roati et~al.}(2008)\citenamefont{Roati, D'Errico,
  Fallani, Fattori, Fort, Zaccanti, Modugno, Modugno, and Inguscio}}]{Inguscio}
\bibinfo{author}{\bibfnamefont{G.}~\bibnamefont{Roati}},
  \bibinfo{author}{\bibfnamefont{C.}~\bibnamefont{D'Errico}},
  \bibinfo{author}{\bibfnamefont{L.}~\bibnamefont{Fallani}},
  \bibinfo{author}{\bibfnamefont{M.}~\bibnamefont{Fattori}},
  \bibinfo{author}{\bibfnamefont{C.}~\bibnamefont{Fort}},
  \bibinfo{author}{\bibfnamefont{M.}~\bibnamefont{Zaccanti}},
  \bibinfo{author}{\bibfnamefont{G.}~\bibnamefont{Modugno}},
  \bibinfo{author}{\bibfnamefont{M.}~\bibnamefont{Modugno}}, \bibnamefont{and}
  \bibinfo{author}{\bibfnamefont{M.}~\bibnamefont{Inguscio}},
  \bibinfo{journal}{Nature} \textbf{\bibinfo{volume}{453}},
  \bibinfo{pages}{895} (\bibinfo{year}{2008}).

\bibitem[{\citenamefont{Keshavamurthy and Schlagheck}(2011)}]{bookS}
\bibinfo{author}{\bibfnamefont{S.}~\bibnamefont{Keshavamurthy}}
  \bibnamefont{and}
  \bibinfo{author}{\bibfnamefont{P.}~\bibnamefont{Schlagheck}},
  \emph{\bibinfo{title}{Dynamical Tunneling: Theory and Experiment}}
  (\bibinfo{publisher}{CRC Press, Singapore}, \bibinfo{year}{2011}).

\end{thebibliography}

\end{document}